\documentclass[hyperref,urlcolor=black]{article}
\usepackage{makeidx}
\usepackage[colorlinks=true,citecolor=black]{hyperref}
\usepackage{hyperref}
\usepackage{multirow}
\usepackage{multicol}
\usepackage[dvipsnames,svgnames,table]{xcolor}
\usepackage{graphicx}
\usepackage{epstopdf}
\usepackage{ulem}
\usepackage{hyperref}
\usepackage{amsmath}
\usepackage{amssymb}
\usepackage{enumerate}
\usepackage{latexsym}
\usepackage[a4paper]{geometry}
\usepackage{caption}
\usepackage{subfigure}
\usepackage{booktabs} %需要加载宏包{booktabs}
\usepackage{url}
\usepackage{verbatim}
\usepackage[justification=centering]{caption} %让所有标题居中
\renewcommand{\thesubfigure}{\arabic{subfigure}} 
\makeatletter
\renewcommand{\@thesubfigure}{(\thesubfigure)\space}
\renewcommand{\p@subfigure}{} 
\makeatother
\hypersetup{
colorlinks=true,
linkcolor=black
}
\begin{document}
\captionsetup[figure]{labelfont={bf},name={Fig.},labelsep=period}
\title{\bf Predicting the Results of LTL Model Checking using Multiple Machine Learning Algorithms}

\date{}
\author{\large Weijun ZHU, Mingliang XU and Jianwei WANG\\
{\sffamily\small School of Information Engineering, Zhengzhou University, Zhengzhou  450001, China}
}
\maketitle
{\noindent\small{\bf Abstract:}
    In this paper, we study how to predict the results of LTL model checking using some machine learning algorithms. Some Kripke structures and LTL formulas and their model checking results are made up data set. The approaches based on the Random Forest (RF), K-Nearest Neighbors (KNN), Decision tree (DT), and Logistic Regression (LR) are used to training and prediction. The experiment results show that the predictive accuracy of the RF, KNN, DT and LR-based approaches are 97.9\%, 98.2\%, 97.1\% and 98.2\%, respectively, as well as the average computation efficiencies of the RF, KNN, DT and LR-based approaches are 7102500, 598, 4132364 and 5543415 times than that of the existing approach, respectively, if the length of each LTL formula is 500.}

\vspace{1ex}

\section{Introduction}
Model checking is a kind of formal verifying technique, which was presented by Turing Award winner Prof. Clarke et al\cite{r1}. This technique is widely used in CPU design\cite{r2}, security protocols \cite{r3} and malware detection\cite{r4} by some leading IT companies, including INTEL and IBM\cite{r5}. And the state explosion problem (In a special case, $10^{120}$ states were verified automatically by the symbolic model checker\cite{r6}.) has been widely concerned. In our previous work, we used the BT algorithm to predict the results of LTL model checking\cite{r7}. Can other machine learning algorithms be used to predict the results of LTL model checking effectively? This is our research issue.

\section{The principle of the new method}
\begin{center}
	\begin{figure}[htb!]
		\centering
		\subfigure[For a given pair of Kripke structure K and a LTL formula f, determine whether K satisfies f or not.]{
			\begin{minipage}[b]{0.55\textwidth}
				\includegraphics[width=\textwidth]{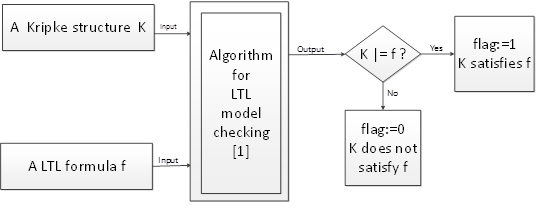} \\
			\end{minipage}
		}
		\subfigure[ The model M can predict the model checking results for $m_2$-$m_1$ pairs of K and f, since M is obtained by training $m_1$ groups of K, f and their model checking result r.]{
			\begin{minipage}[b]{0.55\textwidth}
				\includegraphics[width=\textwidth]{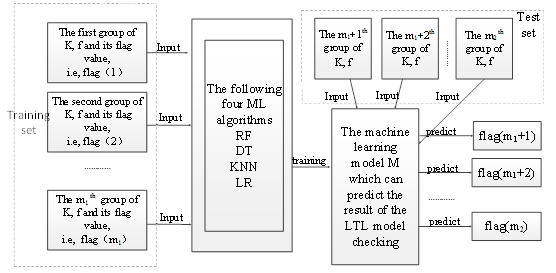} \\
			\end{minipage}
		}
		\caption{Given one pair of model and formula, the new method can determine/predict whether this model satisfies this formula or not. }
		\label{fig1}
	\end{figure}
\end{center}

The principle of the new method can be described as Fig. \ref{fig1}.

\section{Simulated experiments}

\subsection{The simulation platform}
(1)CPU: Intel(R) Core(TM) i7-4790 CPU @3.60GHz.

\noindent
(2)RAM: 8.0G RAM.\\
\noindent
(3)OS: Windows 10.\\
\noindent
(4)NuSMV\cite{r8}: for performing LTL model checking.\\
\noindent
(5)Graph Lab\cite{r9}: for implementing the RF\cite{r10}\cite{r11}, DT\cite{r12}\cite{r13}, KNN\cite{r14}\cite{r15} and LR\cite{r16}\cite{r17} algorithm.

\subsection{Experimental Procedures}
The experimental steps in this paper are consistent with the experimental steps in our previous work, and will not be described repeatedly here. For details, please refer to \cite{r7}.

\subsection{Experimental results }
\subsubsection{NuSMV experiments}
The data sets obtained in this section are raw data for Grap Lab experiments in the next section, where the length of each formula is 25 or 500. The NuSMV experiments in this study is the same as that in our previous work. For details, please refer to \cite{r7}.

\subsubsection{Experiments via Graph Lab}
In the NuSMV experiments, we obtained 405 and 400 records of model checking experimental results via NuSMV, where the length of each formula is 25 and 500, respectively. The two data sets are then exported to Graph Lab to conduct the training and prediction using one of the machine learning algorithms based on RF, KNN, DT, and LR, respectively.

\begin{figure}[htb!]
	\centering
	\subfigure[Predictive result via RF, there are 405 records in sample set.]{
		\begin{minipage}[b]{0.5\textwidth}
			\includegraphics[width=\textwidth]{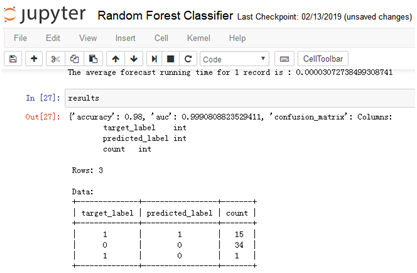} \\
		\end{minipage}
	}
	\subfigure[ Predictive result via DT, there are 405 records in sample set.]{
		\begin{minipage}[b]{0.5\textwidth}
			\includegraphics[width=\textwidth]{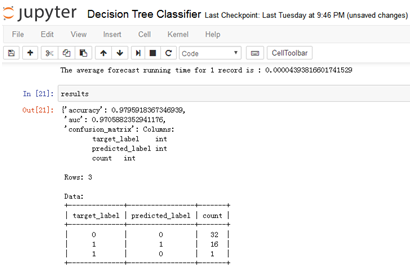} \\
		\end{minipage}
	}
	\subfigure[ Predictive result via KNN, there are 405 records in sample set.]{
	\begin{minipage}[b]{0.5\textwidth}
		\includegraphics[width=\textwidth]{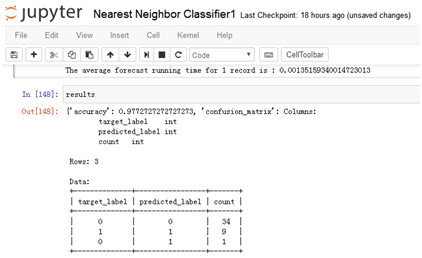} \\
	\end{minipage}
	}
	\subfigure[ Predictive result via LR, there are 405 records in sample set.]{
	\begin{minipage}[b]{0.5\textwidth}
		\includegraphics[width=\textwidth]{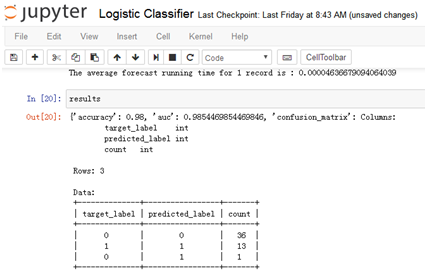} \\
	\end{minipage}
	}
	\caption{Results of prediction based on different ML algorithms, where the length of each formula is 25. }
	\label{fig2}
\end{figure}

The optimal results are summarized in Fig. \ref{fig2} for the four machine learning algorithms, i.e., RF, DT, KNN and LR, where the length of each formula is 25. Table \ref{tab1} shows the value of parameters and some performance when these optimal results occur. Furthermore, the results also indicate that the efficiency of the approaches based on the RF, KNN, DT, LR increase 484, 11, 341, 326 times respectively, compared with the traditional NuSMV model checking approach, as shown in table \ref{tab2}.

\begin{figure}[htb!]
	\centering
	\subfigure[Predictive result via RF, there are 400 records in sample set.]{
		\begin{minipage}[b]{0.48\textwidth}
			\includegraphics[width=\textwidth]{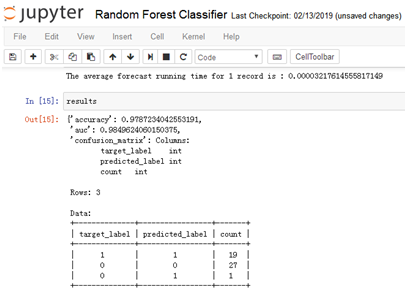} \\
		\end{minipage}
	}
	\subfigure[ Predictive result via DT, there are 400 records in sample set.]{
		\begin{minipage}[b]{0.48\textwidth}
			\includegraphics[width=\textwidth]{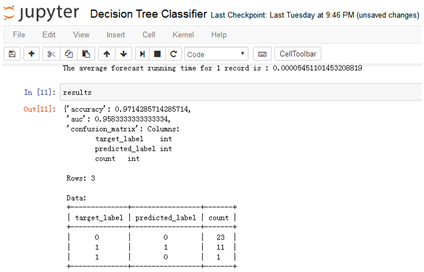} \\
		\end{minipage}
	}
	\subfigure[ Predictive result via KNN, there are 400 records in sample set.]{
		\begin{minipage}[b]{0.48\textwidth}
			\includegraphics[width=\textwidth]{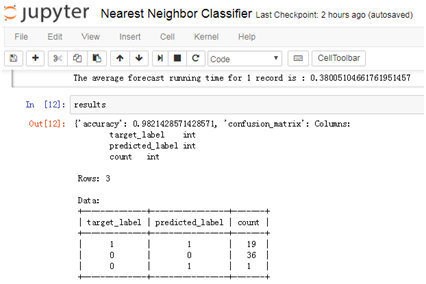} \\
		\end{minipage}
	}
	\subfigure[ Predictive result via LR, there are 400 records in sample set.]{
		\begin{minipage}[b]{0.48\textwidth}
			\includegraphics[width=\textwidth]{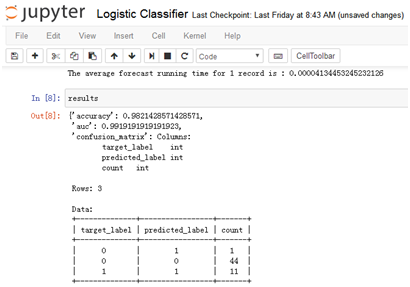} \\
		\end{minipage}
	}
	\caption{Results of prediction based on different ML algorithms, where the length of each formula is 500. }
	\label{fig3}
\end{figure}

The optimal results are summarized in Fig. \ref{fig3} for the four machine learning algorithms, i.e., RF, DT, KNN and LR, where the length of each formula is 500. Table \ref{tab3} shows the value of parameters and some performance when these optimal results occur. Furthermore, the results also indicate that the efficiency of the approaches based on the RF, KNN, DT, LR increase 7102500, 598, 4132364, 5543415 times respectively, compared with the traditional NuSMV model checking approach, as shown in table \ref{tab4}.

\begin{table}[!]
	\centering
	\caption{ML training and testing results in Graph Lab, where the length of each formula is 25}
	\label{tab1}
	\begin{tabular}{ccccc}
		\toprule  %添加表格头部粗线
		ML Algorithms & RF & KNN & DT & LR \\
		\midrule  %添加表格中横线
		Training record \# &	355 &	361 &	356 &	355\\
		Testing record \# &	50 &	44 &	49 &	50\\
		Running time per record (in second) &	 0.000031 & 0.001352 & 0.000044 & 0.000046\\
		Prediction Accuracy &	 0.9800 &	0.9773 & 0.9796 & 0.9800\\
		AUC & 0.9991 &	0.5540 &	0.9706 & 0.9854\\
		Seed \# & 1243 & 1150 & 102 &	691\\
		Fraction &	0.88 & 0.88 & 0.88 &	0.88\\
		\bottomrule %添加表格底部粗线
	\end{tabular}
\end{table}

\begin{table}[!htbp]
	\centering
	\caption{The improvement of the efficiency by our new methods, where the length of each formula is 25}
	\label{tab2}
	\begin{tabular}{cp{4cm}<{\centering}p{4cm}<{\centering}cc}
		\toprule  %添加表格头部粗线
		ML algorithm &	The average time for LTL model checking of one pair of Kripke structure and formula, i.e., $t_1$(in Seconds) &	The average time for predicting the result of model checking, i.e., $t_2$(in Seconds) &	$t_2$/$t_1$ &	$t_1$/$t_2$ \\
		\midrule  %添加表格中横线
		RF &	0.015 &	0.000031 &	0.2\% &	484\\
		KNN &	0.015 &	0.001352 &	9\% &	11\\
		DT &	0.015 &	0.000044 &	0.3\% &	341\\
		LR &	0.015 &	0.000046 &	0.3\% &	326\\
		\bottomrule %添加表格底部粗线
	\end{tabular}
\end{table}

\begin{table}[!htbp]
	\centering
	\caption{ML training and testing results in Graph Lab, where the length of each formula is 500}
	\label{tab3}
	\begin{tabular}{ccccc}
		\toprule  %添加表格头部粗线
		ML Algorithms & RF & KNN & DT & LR \\
		\midrule  %添加表格中横线
		Training record \# &	353 &	344 &	365 &	344\\
		Testing record \# &	47 &	56 &	35 &	56\\
		Running time per record (in second) &	0.000032 &	0.380051 &	0.000055 &	0.000041\\
		Prediction Accuracy & 0.9787 & 0.9821 & 0.9714 & 0.9821\\
		AUC &	0.9850 &0.1387 &0.9583 &0.9919\\
		Seed \# &	858 &	429 &	732 &	2306\\
		Fraction &	0.88 &	0.86 &	0.90 &	0.86\\
		\bottomrule %添加表格底部粗线
	\end{tabular}
\end{table}

\begin{table}[!htbp]
	\centering
	\caption{The improvement of the efficiency by our new methods, where the length of each formula is 500}
	\label{tab4}
	\begin{tabular}{cp{4cm}<{\centering}p{4cm}<{\centering}cc}
		\toprule  %添加表格头部粗线
		ML algorithm &	The average time for LTL model checking of one pair of Kripke structure and formula, i.e., $t_1$(in Seconds) &	The average time for predicting the result of model checking, i.e., $t_2$(in Seconds) &	$t_2$/$t_1$ &	$t_1$/$t_2$ \\
		\midrule  %添加表格中横线
		RF &	227.28 &	0.000032 &	0.000014\% &	7102500\\
		KNN &	227.28 &	0.380051 &	0.2\% &	598\\
		DT &	227.28 &	0.000055 &	0.000024\% &	4132364\\
		LR &	227.28 &	0.000041 &	0.000018\% &	5543415\\
		\bottomrule %添加表格底部粗线
	\end{tabular}
\end{table}

\section{Conclusions}
In this study, we propose a method based on four machine learning algorithms, for predicting the results of LTL model checking. The results show that the longer the formulas involved in the model checking become, the more obvious the comparative advantage of the new method is.

\section*{Acknowledgements}
This work has been supported by the National Natural Science Foundation of China (No.U1204608).

\end{document}